\documentclass{article}
\usepackage{emulateapj}

\usepackage{psfig}

\def\ledd{L_{\rm Edd}}

\def\sw{Schwarzschild~}
\newcommand\fe{Fe K$\alpha$~}

\newcommand\fx{F_{\rm x}}

\newcommand\dm{\dot{m}}
\newcommand\fdisk{F_{\rm d}}

\def\>{$>$}
\def\<{$<$}

\def\simlt{\lower.5ex\hbox{$\; \buildrel < \over \sim \;$}}
\def\simgt{\lower.5ex\hbox{$\; \buildrel > \over \sim \;$}}
\def\sqr#1#2{{\vcenter{\hrule height.#2pt
      \hbox{\vrule width.#2pt height#1pt \kern#1pt
         \vrule width.#2pt}
      \hrule height.#2pt}}}

\begin{document}

\title{Narrow Moving Fe K$\alpha$ lines from magnetic flares in AGN.}

\author{Sergei Nayakshin\altaffilmark{1} \& Demosthenes Kazanas}

\affil{NASA/GSFC, LHEA, Code 661, Greenbelt, MD, 20771}
\altaffiltext{1}{Also Universities Space Research Association}

\begin{abstract}
We point out that luminous magnetic flares, thought to occur in
standard AGN accretion disks, cannot be located much higher than few
pressure scale heights above the disk. Using this fact, we estimate
the fraction of the disk surface illuminated by a typical flare. This
fraction turns out to be very small for geometrically thin disks,
which implies that the instantaneous Fe K$\alpha$ emission line from a
specific magnetic flare is {\em narrow}. The line is red- or
blue-shifted depending on the position of the observer relative to the
flare and sweeps across the line band with time. We present several
examples of theoretical time-resolved line profiles from such flares
for \sw geometry.  The observations of such moving features with
future X-ray telescopes will present a powerful test of the accretion
disk geometry and may also test General Relativity in the strong field
limit.
\end{abstract}

\keywords{accretion, accretion disks ---radiative transfer ---
line: formation --- X-rays: general}

\section{Introduction}\label{sect:intro}

There is much current debate about the structure of the inner part of
the accretion flow in Active Galactic Nuclei (AGN). One of the
possibilities often discussed is the standard (geometrically thin,
optically thick) accretion disk in which the X-ray activity is
provided by magnetic flares above the disk (e.g., Galeev, Rosner \&
Vaiana 1979; Haardt, Maraschi \& Ghisellini 1994). The main difficulty
of this model is its intrinsic physical complexity, which does not
naturally relate the X-ray emission to the {\em global} disk
parameters, such as the accretion rate in the disk, black hole mass
$M$, etc. However the model is attractive because of the physical
parallel with the Solar magnetic flares, and the fact that it
naturally explains the broad \fe lines observed in a number of AGN --
see the recent review by Fabian et al. (2000). Any robust
observational prediction that can be used to test the model is thus of
a substantial value.

Herein we show that time-resolved \fe line profiles of the magnetic
flare model should consist of relatively narrow features due to the
limited extent (in radius and height) of the active regions.  As they
circle the black hole, the corresponding \fe line features should
sweep across $\sim 4-8$ keV energy band. This pattern is unique to the
magnetic flare model (in particular, the line width is much narrower
than the profiles of the lamppost-like models calculated by Reynolds
et al. [1999], Young \& Reynolds [2000] and Ruszkowski [2000]), and it
could be used in the future to verify or falsify the basic premises of
this model.

\section{The flare height and the active region size}

The size of a single magnetic reconnecting region, $\Delta R$, and its
height above an accretion disk are of theoretical and observational
import, yet there is no clear discussion of this issue in the
literature. If one assumes that magnetic field pressure in the flux
tube greatly exceeds the gas pressure there, then its equilibrium
structure should be dictated by a force-free equilibrium.  Parker
(1979, \S 8.4) computes the force-free equilibrium for a flux tube not
bounded by any external pressure. He shows that, mathematically
speaking, the magnetic field fills all the available (empty)
space. However, one can check that most of the magnetic {\em energy}
is confined within a volume of roughly same linear size as the
separation between the footpoints of the tube, i.e., in close
proximity to the footpoints.  Further, it is generally believed that
the size of a magnetic flux tube inside the disk should be no larger
than the size of the largest turbulent cell, which is $\sim H$, the
disk pressure height scale (e.g., Galeev et al.  1979). These
considerations thus require that both the size and the height of a
magnetic flare be $\Delta R \sim H$.

At the same time, one could argue that the magnetic field need not
completely dominate the energy (and pressure) content of the flux tube
above the disk, and hence maybe the gas pressure expands the flux tube
much above $H$. While this is in principle possible dynamically (e.g.,
Romanova et al. 1998), one can show that the corona has to be
magnetically dominated or the energy balance conditions will require
unrealistically large size for the emitting region (e.g., Merloni \&
Fabian 2001). These authors derive also a limit on the size of a
magnetic flare (see their equation 8), which however depends on the
average number of magnetic flares, $N$; this is not a well determined
quantity since the usual variability arguments (e.g., Haardt et
al. 1994) may not be applied if a single flare in the {\em light
curve} is produced by a multitude of reconnecting flux tubes (e.g. as
in the flare avalanche model of Poutanen \& Fabian 1999; also see
below).

We now show that one can derive an even tighter limit on the flux tube
size. The magnetic flux, $\Psi$, is a constant along a flux tube
(e.g., Parker 1979). Since any flux tube is anchored in the disk
mid-plane, the maximum of the magnetic flux is $\Psi \sim B_e H^2$
where $B_e$ is the equipartition magnetic field in the disk mid-plane,
and $H^2$ is the order of magnitude estimate of the maximum tube cross
section within the disk. The magnetic pressure in the loop above the
disk is thus
\begin{equation}
{B^2\over 8\pi}  \sim \beta \lambda^4 P_d \left({H\over \Delta
R}\right)^4\;,
\label{mp}
\end{equation}
where we introduced parameter $\beta \equiv B_{\rm max}^2/B_e^2\leq
1$; $B_{\rm max}$ and $\lambda^2 H^2$ are the actual magnetic field
and the tube's cross section in the mid-plane ($\lambda < 1$), and
$P_d$ is the total disk pressure there.

The constraint on the size of the  active region comes from the demand 
that the mechanism responsible for the formation of the resulting
spectrum be the Comptonization of seed photons by an optically thin, 
mildly relativistic plasma (e.g. Poutanen \& Svensson 1996), a 
process known to yield power law spectra in agreement with those 
observed in AGN.  
%
In order for
Comptonization to be the dominant emission mechanism, the compactness
parameter, $l$, of the emission region must be greater than about 
$\sim 0.1$ (Fabian 1994; Nayakshin 1998, \S 2). To estimate the value 
of $l$ we assume that the luminosity of a flare, $L_1$, is given by
$(4\pi/3) (\Delta R)^3 (B^2/8\pi) \, t_r^{-1}$, where $t_r$ is the
reconnection time scale, here parameterized as $b \, \Delta R/c$ with
$b > 1$. With this assumption, the X-ray flux from the active region 
is $\fx \sim (B^2/8\pi)\,c/3b$, which with the use of Eq.(\ref{mp}) 
yields for the compactness
\begin{equation}
l \equiv {\sigma_T \fx \Delta R\over m_e c^3}= \beta \lambda^{4} {P_d
\sigma_T H\over 3 b m_e c^2}\, \left[{H\over \Delta R}\right]^3\;.
\label{es1}
\end{equation}
It is a simple matter to show, using equations of Svensson \&
Zdziarski (1994; SZ94), that for both radiation- and gas-dominated
disks,
\begin{equation}
{P_d \sigma_T H\over 3 m_e c^2}\, = {1\over 6\sqrt{2}}\,
(m_p/m_e)\alpha^{-1} \eta^{-1} r^{-3/2} \dm J(r)\;,
\label{es2}
\end{equation}
where $\alpha$ is the viscosity parameter; $\dm$ is the dimensionless
accretion rate (for $\dm=1$ the disk luminosity equals that of
Eddington $\ledd$); $\eta\simeq 0.06$ is the efficiency of accretion;
$J(r) \equiv 1 - \sqrt{3/r}$, and $r = R c^2/2 GM$.  The function
$J(r)/r^{3/2}$ has a maximum at $r=16/3$, with the value $\simeq
0.02$. Therefore, for any radius in the disk,
\begin{equation}
l \leq 73\, \beta \lambda^{4} {\dm\over \alpha b} \,\left[{H\over \Delta
R}\right]^3\;.
\label{es3}
\end{equation}
The corresponding limit on the size of
the active region is
\begin{equation}
{\Delta R\over H} \leq 4.2 \left[{l\over
0.1}\right]^{-1/3}\left[{\beta\over \alpha_2 b_1}\right]^{1/3}
\lambda^{4/3} \left({\dm \over 0.01}\right)^{1/3}\;.
\label{es4}
\end{equation}
where $\alpha_2\equiv
\alpha/100$ and $b_1\equiv b/10$.
Thus, the requirement of sufficiently large compactness for an active
region limits its height to no more than several times the disk scale
height $H$. This limit can be even tighter if one demands larger
values for $l$ (e.g. $l\simeq 10$ was assumed by Poutanen \& Fabian
1999).

Let us now compare $H$ with the radius, $R$. Using equation (7) of
SZ94, one finds
\begin{equation}
\frac{H}{R} = \frac{1}{16}\, \frac{\dm}{0.01} \frac{J(r)}{r} \;.
\label{hr}
\end{equation}
This is a tiny number: for $r=5$, i.e., for radii around which the
radiation flux in the Shakura-Sunyaev disk reaches maximum, $H/R
\simeq 0.003$ for $\dm = 0.01$.  Note that our considerations do {\em
not} preclude the presence of larger size magnetic loops, but
they show that these loops will have very weak magnetic fields and
hence are unlikely to be responsible for the X-ray radiation that we
observe from AGN.

Let us now consider the area of the disk illuminated by a flare.  A
source that is a height $\Delta R$ above the disk will illuminate area
$\sim \pi (\Delta R)^2$ immediately below the flare, and the rest of
the disk will receive a negligible amount of X-ray illumination. This
area, $\pi (\Delta R)^2$, is a very small fraction of the disk full
area, according to Eq. (\ref{hr}). This fact has an important
observational implication: the instantaneous \fe line profile from a
single magnetic flare should be {\em narrow} (see below). With an
estimate of the extent $\Delta R$ of a magnetic flare 
(Eq. \ref{es4}) one can now estimate its luminosity as a fraction 
of the total disk luminosity $L = \dm \ledd$ (at $r = 6$):
\begin{equation}
\frac{L_1}{L}\simeq  1.4\times 10^{-2} \frac{\beta
\lambda^4}{\alpha_2 b_1} \left(\frac{H}{\Delta R}\right)^2
\frac{\dm}{0.01}\;.
\label{l1}
\end{equation}
Because both $\lambda$ and $\beta$ must be smaller than unity by
definition, the above relation shows that if the X-ray luminosity is a
good fraction of the bolometric luminosity, then one needs $N$ (the
number of such X-ray emitting regions) $\sim$ from few tens up to few
thousand at any one time. If flares occurred completely randomly,
independently of each other, then one would expect an RMS variability
amplitude $\sim 1/\sqrt{N}$. This latter number seems to be too small
given that it is not atypical for AGN to exhibit variations in X-ray
flux by factors of 2 or so.

To account for the observed variability one needs to invoke avalanches
of magnetic flares, i.e., require that each observed X-ray flare
consists of many correlated (in time and space) individual active
regions (e.g., Poutanen \& Fabian 1999). Physically, it is likely that
magnetic flux tubes rise above the disk but do not immediately
reconnect, settling down into a quasi-static equilibrium
state. However, with time, more and more magnetic flux tubes arise
from the disk, covering the disk surface. Additionally, the tubes that
are already above the disk are sheared by differential rotation of the
foot-points and hence some of them will be taken out of equilibrium
(i.e., the reconnection will start).  If some region of the disk is
particularly closely packed with flux tubes, then it is possible that
the active magnetic flares will affect its neighbors, setting them off
as in a chain reaction, producing the flare avalanche and leading to a
flare-like event in the observer's light curve.

Note that for this scenario to be plausible, the magnetic flares
taking part in the avalanche must be near each other, otherwise it is
hard to see how they can interact over distances greater than their
own size, $\Delta R$. Therefore, from now on, we will accept that the
active flares contributing to a particular observer's flare in the
light curve (which can be as large as $\sim 50$\% of the average X-ray
luminosity of an AGN) closely pack a disk region of area $A$.  Assume
that the active region produces a fraction $\zeta$ of the bolometric
luminosity, $L$. The maximum X-ray flux produced by flares in the
active region is $\fx \sim c P_d b^{-1}$. The area of the disk covered
with magnetic flares responsible for the luminosity $\zeta L$ is thus
$A\sim \zeta L/ \fx$. We can compare this area with the effective area
of the inner accretion disk $\pi R^2$ if we note that $L \sim \fdisk
\pi R^2$, where $\fdisk = (9/8) \alpha c_s P_d$ is the Shakura-Sunyaev
disk flux, and $c_s$ is the mid-plane sound speed:
\begin{equation}
\frac{A}{\pi R^2}\sim \zeta \alpha b \frac{c_s}{c} \simeq 10^{-4}
\alpha \zeta b\;,
\label{ratio}
\end{equation}
where we used $T\sim 10^5$ to estimate $c_s$. The combination of the
remaining factors in the equation above is probably less than unity,
and so clearly $A\ll R^2$.  If we assume that the active region has
about same dimensions in the radial ($\delta R$) and azimuthal (R
$\delta \phi$) directions, then we conclude that $\delta R/R \sim
\delta \phi/2\pi \sim 0.01$.

\section{Time resolved \fe line profile}

An immediate implication of the discussion of the previous section is
the prediction that the instantaneous \fe line profile as seen by an 
observer at
infinity should be narrow. Indeed, for a non-rotating black hole, and
neglecting photon ray bending due to strong gravitational fields for a
moment, the observer sees \fe line photons of energy $E(R, \phi)$ from
a point source at $(R, \phi)$
\begin{equation}
E(R,\phi) = E_0 (1- 2/r)^{1/2} \left[\gamma (1 - v
\cos\alpha_0)\right]^{-1}\;,
\label{e0}
\end{equation}
where $v$ is the orbital velocity in units of speed of light at radius
$r$, $\alpha_0$ is the angle this velocity makes with the direction to
the observer, $\gamma \equiv (1-v^2)^{-1/2}$, and
$E_0$ is the photon rest frame energy.  Approximately, the width of
the line profile will be
\begin{equation}
\delta E = |\frac{\partial E}{\partial R} \delta R | + |\frac{\partial
E}{\partial \phi} \delta \phi |\;.
\label{eshift}
\end{equation}
For a pole-on observer the second term vanishes and $\delta E \sim
0.1 E_0 (\delta R/R) \sim 6$ eV (!) for the chosen values of $\delta R,
\delta \phi$, much
smaller than the red or blue-shift of the line centroid itself.
$\delta E$ generally increases with increasing inclination angle. The
maximum of the second term in eq. (\ref{eshift}) is for an equatorial
observer when $\alpha = \pi/2$ or $3 \pi/2$: max$|\partial E/\partial
\phi| = E(R,\phi) v\simeq E_0 (r-2)^{-1/2}$, which yields $\delta E
\simeq 130 {\rm eV} (\delta \phi/0.01\pi)$. Hence the width of \fe
line profile from a \sw disk flare should vary between as little as
$\sim 6$ eV to $\sim 100$ eV, with a ``reasonable'' value of $\sim 30$
eV for Seyfert 1 Galaxies that are thought to be nearly pole-on.  Note
that the contrast of these features to the continuum flux is
larger (by a factor of $\sim 10$) for nearly face-on than for nearly
edge-on disks.

Let us now calculate the time-resolved \fe line profile from a single
rotating spot. We will assume that its radial size $\delta R$ is
negligible, and $\delta \phi/(2\pi) = 0.01$.  Note that the time delay
between the start of a magnetic flare and the \fe line emission from
the flare is very small, i.e., it is $\sim H/c$, and so we can neglect
it.  For this first study, we will assume that the \fe line emissivity is
isotropic in the rest frame and is also constant over the active region's
life time. Finally, we will consider the case of a flare lasting
exactly one rotation for a non-rotating black hole at a radius $R = 12
GM/c^2$.

In polar spherical coordinates with the black hole at the center and
the disk treated as a plane at $\theta = \pi/2$, we place the observer
at $r_o = 10^3$, azimuthal angle $\phi = 0$ and polar angle $\theta =
\theta_o$. The active region turns on at $\phi = - \pi$ and turns off
at $\phi = \pi$ by assumption. We consider three different 
values of the observer
inclination angles: $\theta_o = 4$, $27$ and $63$ degrees. To
calculate the time dependent line profiles, we isotropically (in the
rest frame of the active region) emit a large number of photons and
then trace their trajectories until they reach radius $r_o$. The
photon ray tracing is performed using appendices A3 and A4 of Reynolds
et al. (1999). Only the photons that arrived within angle $\theta =
\theta_o \pm 2^{\circ}$ and $\phi = \pm 0.005\times 2\pi$ are
recorded. Figure 1 shows the resulting \fe line trajectory in the
energy-time diagram for the three chosen angles. One can easily derive
an analytical expression for these tracks neglecting photon ray
bending (e.g., eq. \ref{eshift} with $\alpha$ calculated as a function
of photon arrival time). This expression agrees with the curves shown
in Figure 1 quite well (because for $r=12$ and not too large
$\theta_o$, the ray bending is relatively weak).

When the observer is nearly pole-on, there is almost no Doppler boost,
so the photon line trajectory in Figure 1 is a straight line . At
larger angles, line photons can be either red or blue shifted
depending on the azimuthal separation of the source and the
observer. Finally, one should note that the larger the inclination
angle, the shorter (in time) the blue-shifted section of the S-shaped
trajectory is. This is simply due to the fact that the source moves
towards the observer for $-\pi < \phi < 0$ and then it moves away from
the observer for $0 < \phi < \pi$.  Also note that similar
trajectories can be noticed in some of the response functions
calculated by Ruszkowski (2000; see his Figures 5,7 \& 9).  The latter
are however much broader because the X-ray source is located much
higher above the disk.

\begin{figure*}[t]
\centerline{\psfig{file=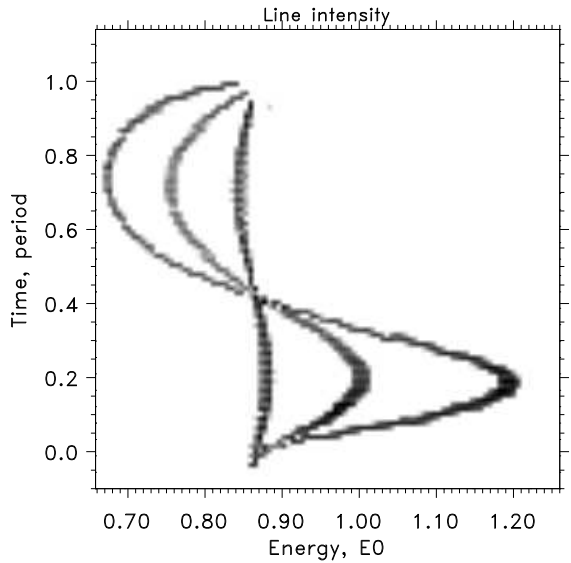,width=.4\textwidth,angle=90}}
\caption{\fe line trajectories in energy-time space for three
different viewing angles: $\theta = 4$, 27 and $63^\circ$. The flare
is located at radius $R = 12 GM/c^2$. The larger the viewing angle,
the larger the spread between maximum and minimum energy observed.}
\label{fig:temp}
\end{figure*}

\section{Discussion}

In this Letter we showed that the area illuminated by X-rays from an
active magnetic flare is small compared to the total disk area.  The
time-resolved \fe line profile from this active region should then be
a narrow feature sweeping with the rotation period across the entire
range of red(blue)-shift associated with the given radius and
inclination angle.  We calculated several examples of such line
profiles for a non-rotating black hole. We now discuss the
implications of our results assuming that such a narrow moving \fe
line trajectory can be observed in the future with observatories such
as XMM-Newton or (more realistically) Constellation-X.

(1) The trajectories of the \fe feature in the energy-time diagram are
very different from those obtained by Reynolds et al. (1999), Young \&
Reynolds (2000) and Ruszkowski (2000) for the lamp-post model or
highly elevated flares.  Indeed, these latter profiles are
considerably broader -- i.e., $\delta E\sim 1$ keV with the red- and
blue-shifted wings of the line appearing simultaneously. This is due
to the fact that a highly elevated X-ray source illuminates a large
fraction of the disk area. In contrast, we found $\delta E \simlt 100$
eV with the line being either red- or blue-shifted at any particular
time. Future observations of sufficient energy and time resolution
should be able to confirm these results thereby setting limits on the
size and height of X-ray emitting regions in accretion disks.

(2) The much more restricted extent of the X-ray emitting region
argued for in this note compared to that of e.g. the lamp-post model
presents a much more unequivocal probe of the underlying geometry. In
the lamp-post case, one of the complications is that even if the
accretion disk proper does not exist at radii smaller than the last
stable orbit, there is still gas there.  This gas is thought to be
free-falling into the black hole.  As shown by Reynolds \& Begelman
(1997), the vertical Thomson depth of this material can be
significant, thus suggesting that it will also produce a fluorescent
\fe line emission if it is illuminated by X-rays. And indeed, if the
X-ray source is located on the height $\simgt$~few~$GM/c^2$ above the
disk, then this region also contributes to the observed line profile
(Reynolds \& Begelman 1997). Young, Ross \& Fabian (1998) argued that
a detailed account of disk photo-ionization can still constrain the
location of the innermost radius of the disk in the particular case of
the well known AGN~~ MCG-6-30-15. For time-dependent reflection, the
distinction between rotating and non-rotating holes is in a red-ward
moving bump in the line profile, as shown by Reynolds et al. (1999)
and Young \& Reynolds (2000). However, recently, it was realized that
thermal ionization instability plays a crucial role in the
photo-ionized reflection (Nayakshin, Kazanas \& Kallman 2000).  
Recent modeling of time-dependent reflection that included thermal
ionization instability (Nayakshin \& Kazanas 2001) shows that the
physics of the problem is far richer than it has been thought based
on constant density models. We therefore feel that distinguishing
between rotating and non-rotating black holes in the lamp-post
geometry can be non trivial in practice.

In contrast, magnetic flares can only occur where there is a
continuous energy production in the underlying disk, that is at radii
greater than the last stable orbit. Since the flares illuminate only
the disk immediately below them, the region inwards of the last stable
orbit receives a negligible amount of X-rays and hence emits no line.
Observation of a \fe line shifted by an amount only consistent with
$R < 6 GM/c^2$, would strongly argue for a rotating black hole, then.

(3) \fe line trajectories similar to those in Figure 1 could be used to
constrain circular orbits of test particles in the gravitational
field, as well as photon ray bending. In principle, this information
can be used to test General Relativity in the strong field limit.
Note that, as long as the line profile is sufficiently narrow, this
procedure would require knowledge of only the \fe line energy versus time
and not the amplitude of the line emission; hence complications due to
anisotropy of the magnetic flare emissivity or photo-ionization
physics may be neglected (this is not so for the lamppost case because
it may be hard to define the peak of the line if the profile is broad
and is overlayed on the top of a continuum emission).

The real \fe line profiles should be produced by many magnetic flares
(see eq. \ref{l1}). If they occur randomly, at completely independent
locations, then it will be perhaps impossible to distinguish
individual line ``trails'' if of sufficiently large number. However, 
as we discussed in \S 2, this would also preclude any substantial
variability in the X-ray continuum flux, while one often
observes variations of up to $\sim 50$ \% in the continuum flux (e.g.,
Edelson et al. 2000). Therefore, we believe that observed large
amplitude excursions of the AGN X-ray flux can only be accounted by
magnetic flares if the latter are not independent of each other, i.e.,
if they take part in flare avalanches (Poutanen \& Fabian
1999). Further, it is difficult to see how small scale flares can
affect each other unless they are in physical proximity. Therefore, 
the line profile from a magnetic flare avalanche should also be narrow
and hence the points made above hold true. On the other hand, the
active region (the avalanche region) may be sheared by the
differential rotation in the disk (although magnetic field can also be
strong enough throughout the active region to inhibit differential
rotation [E. Vishniac, private communication]). Thus, the region may
become larger in extent and the point-source approximation will be
invalid.  Whether this is the case or not could be determined directly
from time-resolved \fe line spectroscopy.


In summary, we conclude that instantaneous \fe line profiles from
magnetic flares and highly elevated X-ray sources, such as a lamppost,
are very much different (e.g., compare Fig. 1 to Figures in Reynolds
et al. 1999), and hence time-resolved \fe line observations have an
enormous potential for constraining accretion disk theories.  Given
this, Astrophysics community should spare no effort in developing
future space observatories such as Constellation-X.

{}

\end{document}